\documentclass[pdflatex,sn-mathphys-num]{sn-jnl}

\usepackage{graphicx}%
\usepackage{multirow}%
\usepackage{amsmath,amssymb,amsfonts}%
\usepackage{amsthm}%
\usepackage{mathrsfs}%
\usepackage[title]{appendix}%
\usepackage{xcolor}%
\usepackage{textcomp}%
\usepackage{manyfoot}%
\usepackage{booktabs}%
\usepackage{algorithm}%
\usepackage{algorithmicx}%
\usepackage{algpseudocode}%
\usepackage{listings}%

\theoremstyle{thmstyleone}%
\theoremstyle{thmstyletwo}%

\theoremstyle{thmstylethree}%

\begin{document}

\title[Article Title]{The LHCb Stripping Project: Sustainable Legacy Data Processing for High-Energy Physics}


\author*[1]{\fnm{Nathan Allen} \sur{Grieser}}\email{ngrieser@cern.ch}

\author[2]{\fnm{Eduardo} \sur{Rodrigues}}\email{ eduardo.rodrigues@liverpool.ac.uk}

\author*[3]{\fnm{Niladri} \sur{Sahoo}}\email{niladri.sahoo@cern.ch}

\author*[4,5]{\fnm{Shuqi} \sur{Sheng}}\email{shuqi.sheng@cern.ch}

\author[6]{\fnm{Nicole} \sur{Skidmore}}\email{nicola.skidmore@cern.ch}

\author[7]{\fnm{Mark} \sur{Smith}}\email{mark.smith@cern.ch}

\affil[1]{University of Cincinnati, Cincinnati, United States}

\affil[2]{Oliver Lodge Laboratory, University of Liverpool, Liverpool, United Kingdom}

\affil[3]{University of Birmingham, Birmingham, United Kingdom}

\affil[4]{École Polytechnique Fédérale de Lausanne, Lausanne, Switzerland}

\affil[5]{Institute of High Energy Physics, Chinese Academy of Sciences, Beijing, China}

\affil[6]{Department of Physics, University of Warwick, Coventry, United Kingdom}

\affil[7]{Imperial College, London, United Kingdom}

\abstract{

The LHCb Stripping project is a pivotal component of the experiment’s data processing framework, designed to refine vast volumes of collision data into manageable samples for offline analysis. It ensures the re-analysis of Runs 1 and 2 legacy data, maintains the software stack, and executes (re-)Stripping campaigns. As the focus shifts toward newer data sets, the project continues to optimize infrastructure for both legacy and live data processing. This paper provides a comprehensive overview of the Stripping framework, detailing its Python-configurable architecture, integration with LHCb computing systems, and large-scale campaign management. We highlight organizational advancements such as GitLab-based workflows, continuous integration, automation, and parallelized processing, alongside computational challenges. Finally, we discuss lessons learned and outline a future road-map to sustain efficient access to valuable physics legacy data sets for the LHCb collaboration.
}


\maketitle

\section{Introduction}
\label{Intro}

The LHCb Stripping project~\cite{Stripping_Git} plays a vital role in filtering the experiment's data, serving as the last central offline processing step where physicists select interesting particle interactions from billions of recorded events. Its flexible Python interface allows researchers to customize selection criteria for different physics studies.

Integrated into the LHCb Data Processing \& Analysis (DPA)~\cite{Abdelmotteleb2025} framework since 2020, the Stripping project is a central pillar of DPA's Work Package 5 (WP5)~\cite{Skidmore:2022rza} , which focuses on legacy software and data. WP5 ensures that legacy data collected at CERN's Large Hadron Collider (LHC)~\cite{Lyndon_Evans_2008, Brüning:782076} during Runs 1 (2010-2012)  and 2 (2015-2018) remain accessible for future re-analysis by maintaining the legacy software stack and organizing necessary (re-)Stripping campaigns. These campaigns, which ran concurrently with data-taking or during End-of-Year (EoY) periods, involve skimming and slimming data to extract the most relevant information for physics analysis.

The Stripping project is designed to streamline the raw data into a manageable subset for offline analysis that maximize signal retention while minimizing background. Its Python-configurable architecture allows for flexible and user-oriented data selection, while modern organizational tools, such as GitLab Milestones, are employed to track developments and ensure timely completion of tasks. 

\section{LHCb Dataflow and Stripping Software Framework}
\label{Stripping}
The LHCb data processing model spans multiple stages, from initial triggering to offline data selection. Stripping reduces the volume of recorded data by selecting events of interest and saving only the necessary information for physics analysis, producing smaller, more manageable data sets. The Stripping software stack~\cite{Stripping_Git} integrates closely with the \texttt{DaVinci}~\cite{DV} framework, which provides essential analysis tools.

\begin{figure}[htbp]
    \centering
    \includegraphics[width=1.0\textwidth]{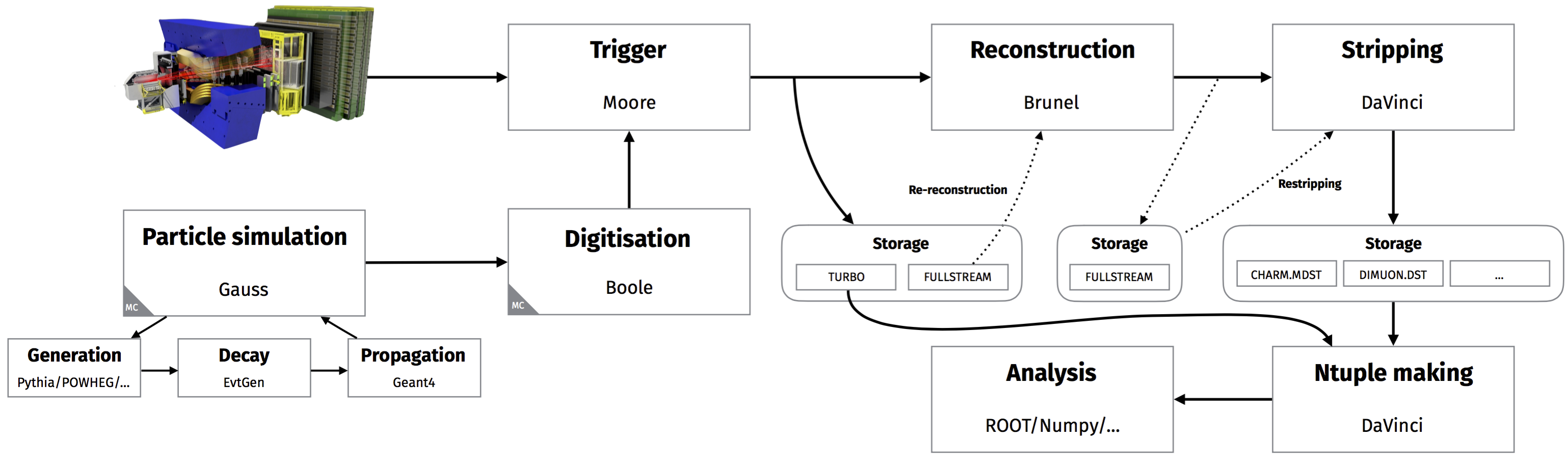}
    \caption{The LHCb dataflow in Run 2~\cite{CERN-LHCC-2018-007} as described in Section~\ref{Stripping}.  The (re-)Stripping stage serves as the last centralized production stage before offline analysis workflows.}
    \label{fig:lhcb_run_2_data_flow}
\end{figure}

The LHC provides proton-proton collisions to the LHCb experiment at an interaction rate of 40 MHz, generating an enormous volume of data--approximately 1 TB every second in Run 2. Storing all this data is impractical due to cost constraints, necessitating a sophisticated filtering process to retain only the most interesting events as shown in Fig.~\ref{fig:lhcb_run_2_data_flow}. This raises critical challenges: how to process and filter data quickly and accurately, manage complex tasks, organize collision data flexibly, and configure software without recompilation. These challenges stem from the high  data arrival rate and the complexity of the data, which contains thousands of potential decay combinations of interest. To address these issues, the LHCb experiment employs a multi-step data processing chain. The Run 1+2 data flow begins with the trigger system, which filters events using hardware (Level-0, L0) and software (High Level Trigger, HLT) components~\cite{The_LHCb_Collaboration_2008}. 
The triggered data are then reconstructed by the \texttt{Brunel} application~\cite{Brunel}, transforming raw detector hits into tracks and particle identification information stored in DST (Data Summary Tape)~\footnote{Files resulting from the reconstruction of real data and of MC samples, basically a ROOT file with our data in an optimized format} files, allowing candidates selected by the HLT to be saved directly to disk. Further filtering is performed through the Stripping process, managed by the \texttt{DaVinci} application, which outputs data in DST or µDST (micro-DST) formats, grouped into physics-specific streams to optimize storage and analysis efficiency.

Simulated data are processed through an identical reconstruction chain as real data, maintaining consistent treatment of detector effects and systematic uncertainties.

The simulation begins with the Gauss~\cite{Clemencic_2011} application, which models proton-proton collisions and particle decays using generators like Pythia~\cite{Sj_strand_2006} and EvtGen~\cite{LANGE2001152}, followed by Geant4~\cite{AGOSTINELLI2003250} for detector response simulation. The Boole~\cite{Clemencic_2011} application converts simulated hits into signals mimicking real detector output, allowing the simulated data to be processed through the same reconstruction and Stripping steps as data. 

The LHCb software is modular, with each application (e.g., \texttt{Brunel}, \texttt{DaVinci}) handling specific tasks, enabling flexibility and efficiency in data processing. Stripping campaigns, identified by versions like SXrYpZ, are central to defining the available reconstructed decays for analysis. Major Stripping versions (X) represent full reprocessing and correspond to data set years.  Minor versions (Y) correspond to superseding full reprocessings, while patch versions (Z) indicate incremental updates. Understanding the reconstruction and Stripping versions is crucial for selecting data, as these versions significantly impact the physics results. The Stripping project website provides detailed information on Stripping configurations, algorithms, and selection criteria, serving as essential resources for analysts.

The Stripping project is maintained across multiple Git production branches, each tailored to specific data-taking periods and campaign types. These branches are designed to support legacy data processing while incorporating necessary bug fixes and updates. New releases are prepared following the completion of development and validation phases, facilitated by an in-depth testing infrastructure. 

\begin{figure}[htbp]
    \centering
    \includegraphics[width=0.5\textwidth]{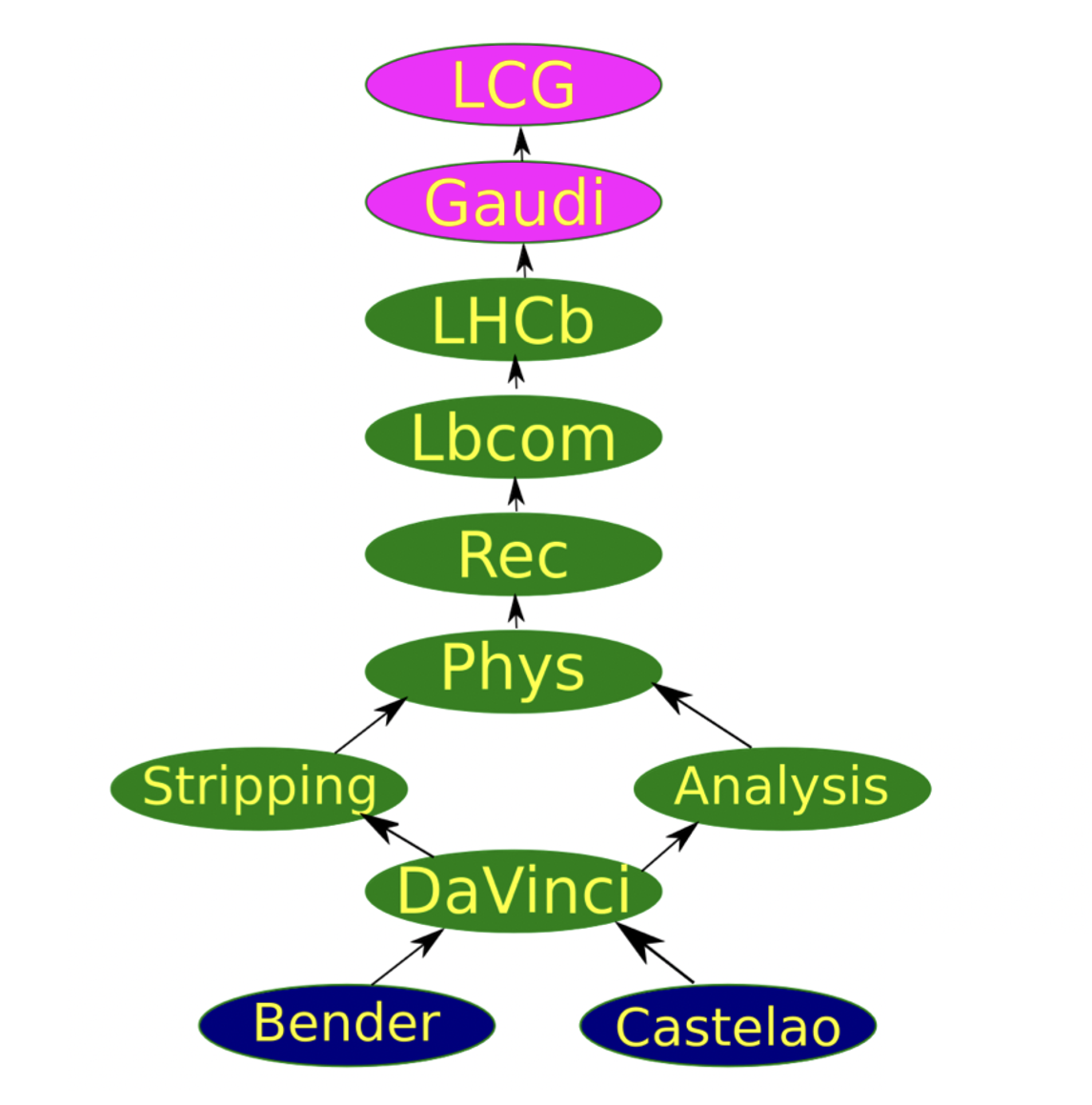}
    \caption{Schematics of the legacy stack utilized for the LHCb offline data processing stage during Runs 1 and 2.}
    \label{fig:software_flow}
\end{figure}

The LHCb Stripping software project is operating atop the primary analysis software, \texttt{DaVinci}, which is built on the \texttt{Gaudi}~\cite{Barrand:2001ny} and \texttt{LHCb} frameworks. The Stripping package, which defines all selection lines and their configurations, is part of a larger software stack, as shown in Fig.~\ref{fig:software_flow}, that includes dependencies such as \texttt{Phys}~\cite{Phys}, \texttt{Rec}~\cite{Rec}, \texttt{LbCom}~\cite{Lbcom}, and \texttt{LHCb}~\cite{LHCb}. This stack relies on CERN's LCG software distribution~\cite{LCG}, with specific ROOT versions determined by the LCG release.

The LHCb Stripping project performs the crucial first stage of offline data processing, transforming reconstructed collision data into manageable subsets for physics analysis. Using a Python-configurable architecture, it enables analysts to select specific physics candidates from the processed datasets, which are stored as 5 GB files on both disk and tape (with placement optimized for storage availability). When new physics selections are developed or existing ones require improved performance, the collaboration conducts systematic ``(Re-)Stripping campaigns" -- comprehensive reprocessing efforts that apply updated selection criteria to the full legacy datasets, ensuring consistent analysis selections across LHCb's physics program.

Stripping campaigns are large-scale efforts that involve collaboration-wide discussions, preparation of selection lines by analysts and Physics Working Groups (PWGs), software stack preparation, validation through mini-productions, and the final production of data sets. These campaigns are categorized into three types: prompt campaigns during data-taking (now obsolete), full campaigns, and incremental campaigns. Full campaigns process all available selection lines, while incremental campaigns focus on newly added or modified lines. 

The campaigns are coordinated by Stripping coordinators, PWG liaisons, analysts, and the Distributed Computing team, with planning and execution spanning several months. The Stripping project's evolution has seen the consolidation of legacy production branches to reduce maintenance overhead. The project's structure and workflow ensure the preservation of legacy data while supporting ongoing and future physics analyses, with detailed documentation and release schedules provided to facilitate collaboration-wide coordination.

\section{Stripping Campaign Workflow and Organization}

The streamlined workflow outlined in this section represents the optimized framework refined during the 2023–2024 incremental re-Stripping campaign, incorporating significant enhancements over previous iterations. This current framework establishes a flexible baseline for future campaigns, facilitating integration of modern and evolving toolkits.

This reprocessing initiative was launched in response to analysis requirements. A pre-campaign needs assessment revealed that most Physics Working Groups (PWGs) primarily aimed to improve ongoing analyses through updated selection criteria.

\subsection{Roles and Responsibilities in the Stripping Campaign}

\begin{figure}[htbp]
    \centering
    \includegraphics[width=1.0\textwidth]{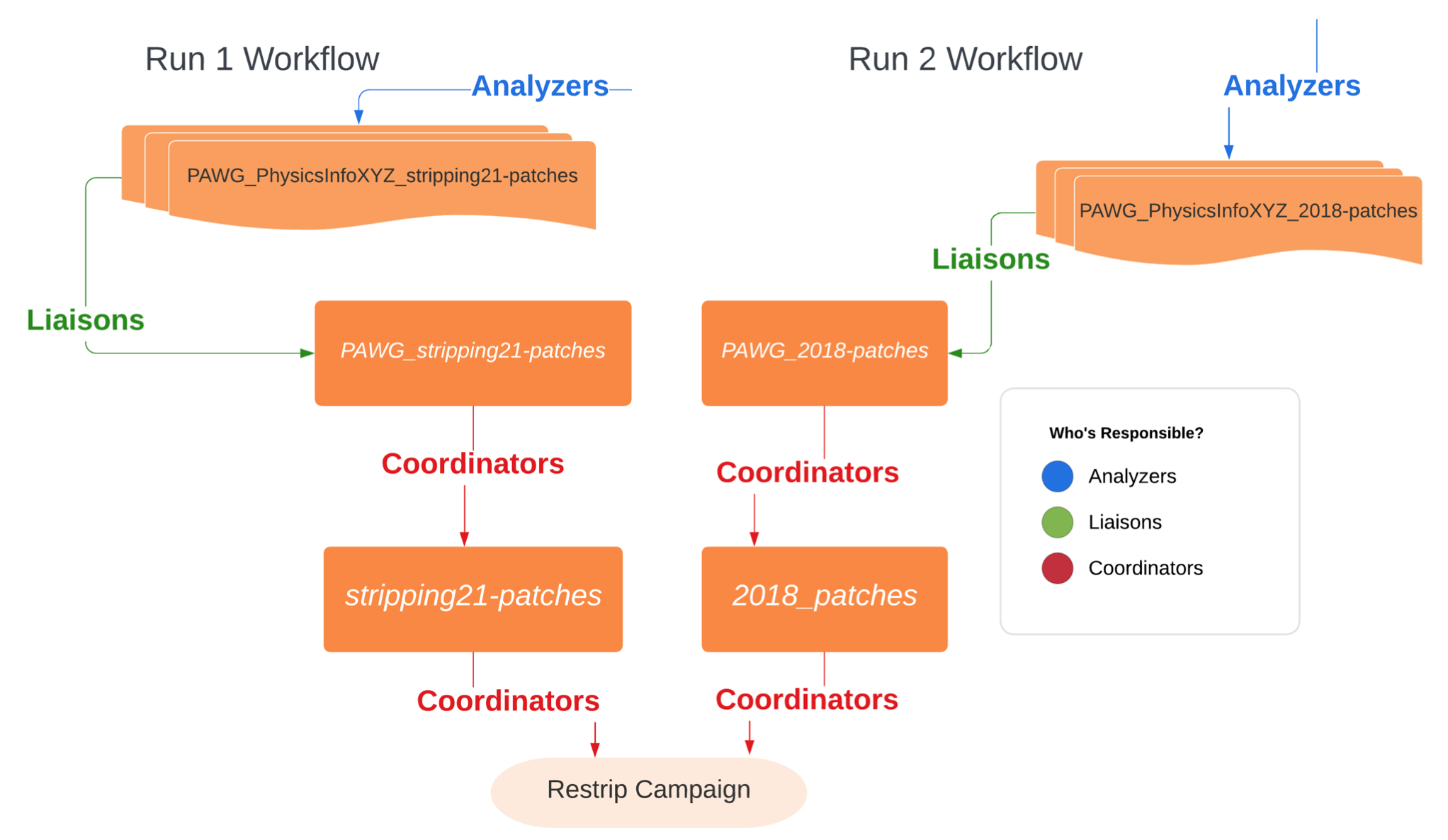}
    \caption{GitLab workflow for the development of the recent Stripping campaign.  The workflow is compartmentalized to allow significant reductions of top-level review, while allowing lower level reviews to have a closer focus on the physics performance.}
    \label{fig:gitlab_flow}
\end{figure}

The central Stripping coordination team serve as the interface between Operations and Physics Planning Groups. Their responsibilities encompassed preparing testing infrastructure (repositories, data files, and \texttt{DaVinci} caches), establishing validation procedures, reviewing and accepted GitLab Merge Requests (MRs)~\footnote{Developers submit an MR to request that their changes be reviewed and merged into the target branch.}, implementing nightly tests for \texttt{DaVinci} and \texttt{Stripping} projects, and ultimately managing production requests. They also maintained campaign documentation and workflow oversight, as illustrated in Figure~\ref{fig:gitlab_flow}.

Each PWG appoints liaisons to facilitate the Stripping campaign. Their key duties included disseminating critical Stripping information (deadlines, workflows, documentation) to their PWG members, assisting with line writing and validation, and monitoring merge requests (MRs) to ensure successful Continuous Integration(CI) tests, an automated process that runs tests on code changes whenever they are pushed to a GitLab repository. Liaisons were specifically responsible for testing in two core packages: (1) \texttt{Phys/StrippingSelections}, verifying line rates and timing constraints, and (2) \texttt{Phys/StrippingSettings}, preparing and validating PWG-wide configuration files before developer handoff.

\subsection{Stripping Campaign Methodology}

The re-Stripping campaign follows a well-defined workflow comprising three key phases: line development, validation testing, and final production approval. While these campaigns have traditionally operated as large-scale coordinated efforts with strict timelines, the latest campaign marked a significant transition in execution procedure, with significant changes to the workflow and management procedures.

\subsubsection{Development, Code Review and Merge Request Workflow}

During the Stripping campaign, analysts develop modules in their respective packages and submit MRs to the PWG-specific development branch. Liaisons oversee these MRs by: (1) ensuring proper naming conventions, correct branch assignments, and linkage to relevant GitLab Milestones and Issues for tracking; (2) validating that Stripping lines meet performance thresholds ($\leqslant$0.05\% rate for DST, $\leqslant$0.5\% for µDST, and $\leqslant$ 1 ms per event timing); and (3) confirming CI tests pass before merging. GitLab's CI system ensures code compatibility by running functional tests on physics selections, monitoring algorithm rates and timing, and organizing PWG contributions efficiently. These CI tests perform linting and execute dataset-specific tests for each year's data. Additionally, to reduce operational overhead, CI tests are defined to only run specific working groups' lines on the related development branches. After approval, liaisons prepare line dictionaries, enabling flexible Stripping campaign configuration per version, for final review and ensure all PWG branches are merged into the production branch by the following week to maintain the validation schedule.  

\subsubsection{Validation Phase and Performance Verification}

Following the development stage, a dedicated testing and validation phase, conducted in close collaboration with PWG liaisons, ensures selections and physics performance meet expectations. Analysts use larger datasets, produced using the official, central production workflow, to identify bugs or unintended selection effects, following the same workflow as the development stage. To maintain the schedule, no new selection lines may be introduced during validation.

\subsubsection{YAML-based Production Management}
The Stripping production workflow has been modernized through the transition from JIRA~\cite{JIRA} to GitLab for coordinator--production team communication, in alignment with Run-3 offline production standards. The new workflow implements YAML configuration files to manage Dirac~\cite{DIRAC} requests, containing complete processing parameters (including years, conditions, data packages and applications) for improved re-usability. Notably, when a campaign's validation phase is completed ahead of schedule,  opportunities to further refine the production request process are able to be included in the overall campaign workflow.

\subsection{Event Metrics in Stripping Campaigns}

This section presents a detailed analysis of Stripped events, comparing per-stream and globally averaged metrics across recent and past re-Stripping campaigns.

The latest incremental re-Stripping of Run 1 (2.5 PB input) reduced the output to 350 TB, achieving a size reduction factor of 7. A 2023–2024 incremental campaign on 6.17 PB of 2016–2018 data produced 1.48 PB, reducing the total size by a factor of 4.2. Further details on event counts, files, and stream-specific storage allocations are provided in Table~\ref{table:latest_stripping_campaigns}.
For Run 2, incremental campaigns consistently yield smaller outputs (by a factor of $\geqslant$2) compared to full re-Stripping due to selective processing.

\begin{table}[htbp]
\centering
\caption{Sample sizes in TB for the latest incremental re-Stripping campaigns (except the 2015 data set, which has not received an incremental re-Stripping following its latest full re-Stripping). The "Processing" is the version of campaign. The "RAW" and "out" columns refer to the input and output samples, both in units of TB.}
\begin{tabular}{c|c|c|c|c}
\hline
Year &Processing & total size (RAW) & size out & reduction factor \\
\hline
2011 & Stripping21r1p2 & 644.0  & 78.2 & 8.2\\
\hline
2012 & Stripping21r0p2 & 1795.2 & 271.3 & 6.6\\
\hline
2015 & Stripping24r2 & 882.2 & 161.1 &5.5\\
\hline
2016 & Stripping28r2p2 & 2657.0 & 476.7&5.6\\
\hline
2017 & Stripping29r2p3 & 2276.0 & 444.6&5.1\\
\hline
2018 & Stripping34r0p3 & 2754.0 & 557.8&4.9\\
\hline
\end{tabular}
\label{table:latest_stripping_campaigns}
\end{table}

Table~\ref{table:latest_stripping_campaigns_rate} provides a yearly breakdown, categorized by magnet polarity, of the Stripping campaigns summarized in Table~\ref{table:latest_stripping_campaigns}.
It details the average per-event sizes of the input and output samples as well as the reduction factor in the total number
(The results were averaged across all streams, with each stream specifically optimized for different physics analyses as outlined in the following.)
of Stripped events.
There is broad agreement of all quantities irrespective of the magnet polarity. Events selected in the 2012 samples are 10\% larger than events selected in the 2011 samples, which reflects a higher event multiplicity given the larger beam energy in 2012 (4.0~TeV) compared to 2011 (3.5~TeV). 


\begin{table}[htbp]
\centering
\caption{Average per-event sizes in kB for the latest re-Stripping campaigns.  The ``Reduction rate'' indicates the reduction in the total size of the samples Stripped.}
\begin{tabular}{c|c|c|c|c|c}
\hline
Year & Processing & Polarity & Avg kB/evt (in) & Avg kB/evt (out) & Reduction rate \\
\hline
\multirow{2}{*}{2011} & \multirow{2}{*}{Stripping21r1p2} &Down   & 58.3 &30.2 & 7.9 \\
&  &Up   & 58.3 &30.1 & 8.7 \\
\hline
\multirow{2}{*}{2012} & \multirow{2}{*}{Stripping21r0p2} &Down  & 64.0 &34.0 & 6.8 \\
&  &Up  & 65.0 &34.0 & 6.5 \\
\hline
\multirow{2}{*}{2015} & \multirow{2}{*}{Stripping24r2} &Down  & 39.0 &29.7 & 3.9 \\
&  &Up  &37.1 &29.9 & 4.0 \\
\hline
\multirow{2}{*}{2016} & \multirow{2}{*}{Stripping28r2p2} &Down  & 49.6 &32.3 & 4.3 \\
&  &Up  &48.1 &32.7 & 4.3 \\
\hline
\multirow{2}{*}{2017} & \multirow{2}{*}{Stripping29r2p3} &Down  & 53.9 &31.2 & 4.2 \\
&  &Up  &56.2 &31.2 & 4.4 \\
\hline
\multirow{2}{*}{2018} & \multirow{2}{*}{Stripping34r0p3} &Down & 55.8 &31.2 & 3.9 \\
&  &Up  &55.3 &31.0 & 4.0 \\
\hline
\end{tabular}
\label{table:latest_stripping_campaigns_rate}
\end{table}

For all Run-2 campaigns considered, similar conclusions can be made, though the average event sizes are more homogeneous in a given campaign --- as can be seen by comparing for example the latest campaign Stripping34r0p3 (2018), Stripping29r2p3 (2017) and Stripping28r2p2 (2016) --- given that the beam energy was always 6.5~TeV (for proton--proton ($pp$) collisions).
Average (over all output streams) Run-2 Stripped event sizes are around 25--35~kB, similar to the event sizes for Run 1.
Large differences in the ratio of events selected by the Stripping and the total sample size reduction are observed among years but most significantly between incremental and full re-Stripping campaigns, again according to expectations due to trigger evolution during Run 2.
For full campaigns the fraction of selected events (over all streams) is often as high as 70\%, a number that seems rather large. This illustrates that slimming events --- rather than skimming events --- can result in larger reductions between tape and disk.
Typical total sample size reductions for incremental and full re-Stripping campaigns are 4--5 and 2--3, respectively.

Output streams are systematically defined for both Run-1 and Run-2 campaigns as below:

\begin{enumerate} 
\setlength{\itemindent}{2em} 
    \item \texttt{BHADRON.MDST}
    \item \texttt{BHADRONCOMPLETEEVENT.DST} 
    \item \texttt{CHARM.MDST}
    \item \texttt{CHARMCOMPLETEEVENT.DST}
    \item \texttt{DIMUON.DST}
    \item \texttt{EW.DST}
    \item \texttt{LEPTONIC.MDST}
    \item \texttt{SEMILEPTONIC.DST}
\end{enumerate}

The majority of campaigns utilized between 7 and 9 distinct output streams, with many specifically designed for individual PWGs---such as \texttt{EW.DST} and \texttt{SEMILEPTONIC.DST}---while others accommodated more comprehensive physics analyses involving multiple PWGs. Specialized streams include \texttt{CALIBRATION.DST} for detector calibration and \texttt{FTAG.DST} for flavor tagging.

\begin{figure}[htbp]
    \centering
    \includegraphics[width=1.0\textwidth]{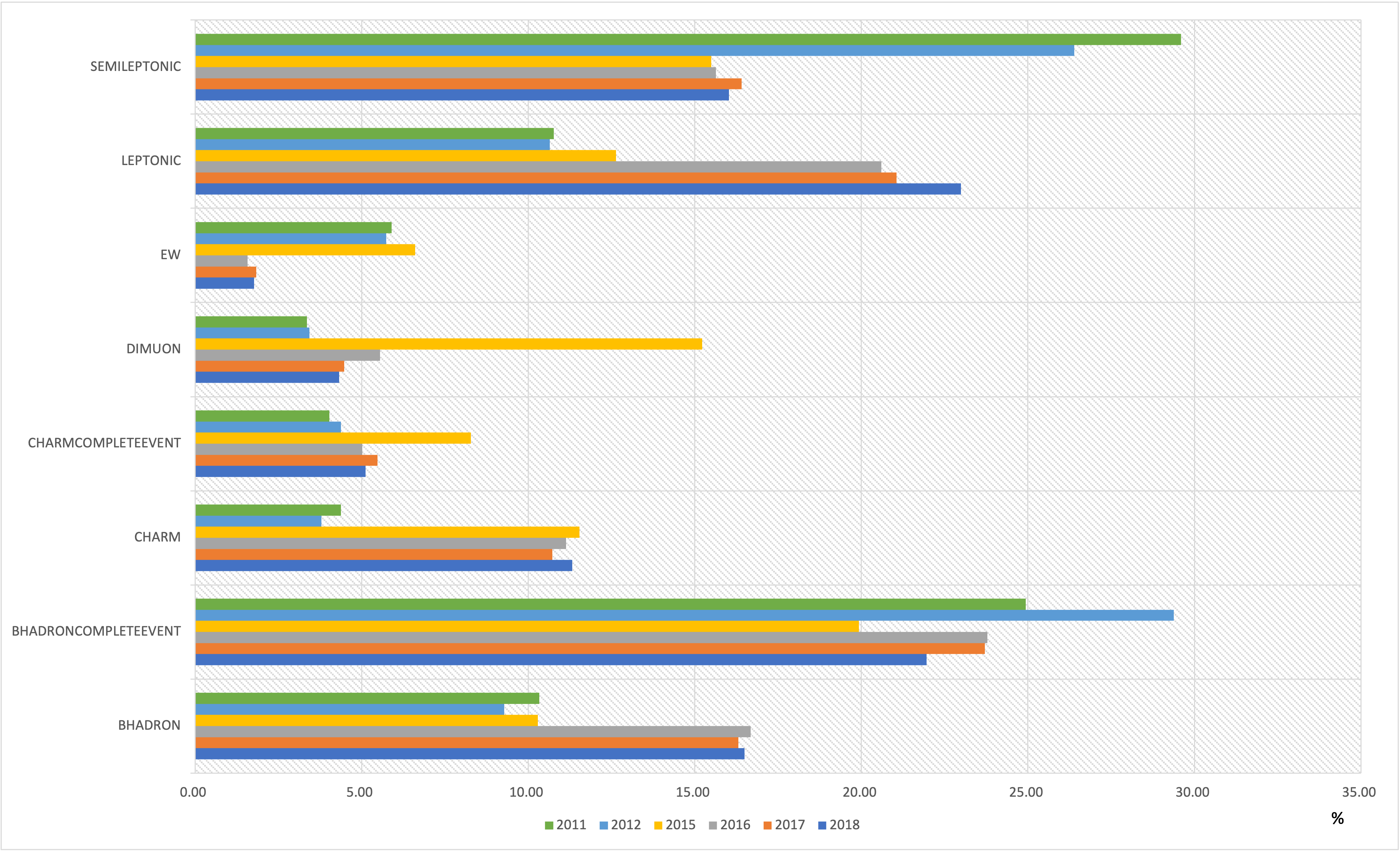}
    \caption{Share of storage space taken by the various output streams in Table~\ref{table:latest_stripping_campaigns}.  Values are provided as percentages out of 100.}
\label{fig:stripping_pp_streams_rel_sizes}
\end{figure}

Fig.~\ref{fig:stripping_pp_streams_rel_sizes} illustrates the relative storage space allocation across different output streams. The observed distribution of output sample sizes for the latest Run-1 and Run-2 incremental campaigns (detailed in Table~\ref{table:latest_stripping_campaigns}) is consistent with expectations, as essentially identical selection criteria were applied across all campaign years. Notably, the \texttt{BHADRONCOMPLETEEVENT.DST}, \texttt{LEPTONIC.MDST}, and \texttt{SEMILEPTONIC.DST} streams collectively account for approximately two-thirds of the total output volume.

\begin{figure}[htbp]
    \centering
    \includegraphics[width=1.0\textwidth]{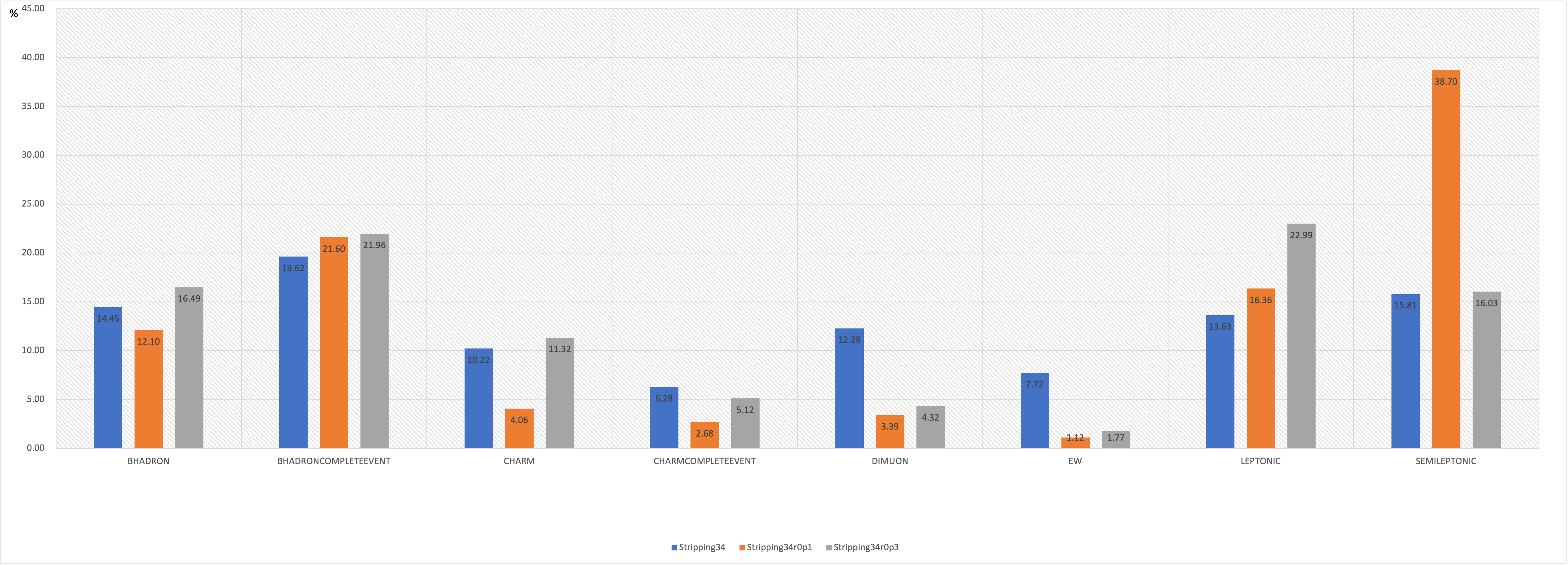}
    \caption{Share of storage space taken by the various output streams in two Run-2 2018 campaigns.  Values are provided as percentages out of 100.}
\label{fig:full_incremental_compare_2018}
\end{figure}

For the 2018 dataset, Fig.~\ref{fig:full_incremental_compare_2018} shows the comparison between the full Stripping34 campaign and various incremental re-Stripping campaigns. Large differences can be observed in the shares taken by the various output streams. Though a detailed comparison is non-trivial and of limited interest, it is a fact that the differences directly reflect the sets of selections included by the various PWGs in any given campaign.

Table~\ref{table:Dimuon_campaigns_rate} details the number of Stripped events and, most importantly, the total output sizes (in TB) and average per-event sizes (in kB) for the \texttt{DIMUON.DST} stream that was produced by any of the considered campaigns.
Wide differences in average event size are seen across the streams.
For example, a typical event size in the DST output streams (rather than µDST) is 80--100~kB --- here, streams persist selected events in full, which analysts can further process downstream. In comparison, events in ``standard" streams, e.g. the \texttt{BHADRON.MDST} stream, require as little as 10--15~kB,
though average sizes up to 50~kB are also seen,
in particular for full campaigns.
The \texttt{SEMILEPTONIC.DST} stream, whereby partially reconstructed events must be selected, lies somewhat in the middle, with average event sizes around 70--80~kB.
The average event sizes depend little on whether a campaign is incremental or not, as expected, since the average per-event size should not depend much on the set of lines run.

\begin{table}[htbp]
\centering
\caption{Number of Stripped events, total storage sizes (in TB) and average per-event sizes for the \texttt{DIMUON.DST} stream for all campaigns compared.}
\begin{tabular}{c|c|c|c|c|c}
\hline
Year & Processing & Polarity & events (out) &Total Size (out)  & Avg kB/evt (out) \\
\hline
\multirow{2}{*}{2011} & \multirow{2}{*}{Stripping21r1p2} &Down   & 19,674,496 &1.6 & 80.0 \\
&  &Up   & 13,437,108 &1.1 & 78.7 \\
\hline
\multirow{2}{*}{2012} & \multirow{2}{*}{Stripping21r0p2} &Down  & 49,560,939 &4.7 & 93.9 \\
&  &Up  & 49,268,945 &4.6 & 94.2 \\
\hline
\multirow{2}{*}{2015} & \multirow{2}{*}{Stripping24r2} &Down  & 132,938,710 &14.2 & 107.1 \\
&  &Up  &96,773,326 &10.3 & 106.2 \\
\hline
\multirow{2}{*}{2016} & \multirow{2}{*}{Stripping28r2p2} &Down  & 105,990,750 &13.7 & 129.1 \\
&  &Up  &100,173,860 &12.8 & 127.6 \\
\hline
\multirow{2}{*}{2017} & \multirow{2}{*}{Stripping29r2p3} &Down  &88,091,847 &10.3 & 117.3 \\
&  &Up  &82,218,042 &9.6 & 116.5 \\
\hline
\multirow{2}{*}{2018} & \multirow{2}{*}{Stripping34r0p3} &Down & 100,096,435 &11.9 & 119.3 \\
&  &Up  &104,280,590 &12.2 & 116.5 \\
\hline
\end{tabular}
\label{table:Dimuon_campaigns_rate}
\end{table}

Analysis of the SMOG\footnote{The LHCb System for Measuring the Overlap with Gas (SMOG) system enabled fixed-target proton-noble gas collisions during Run 2 by injecting a precise density of gas into the beam pipe intersection region.}~\cite{SMOG} heavy-ion data reveals atypical event size evolution, where final \texttt{IFT\footnote{IFT refers to the Ion and Fixed Target PWG of LHCb}.DST} outputs (35~kB/event) triple intermediate-stage sizes, contrasting with $pp$ campaigns, while Stripping maintains 110-130~TB total volume through a factor 3 reduction.

\section{Operational and computing aspects}

\begin{figure}[htbp]
    \centering
    \includegraphics[width=0.9\textwidth]{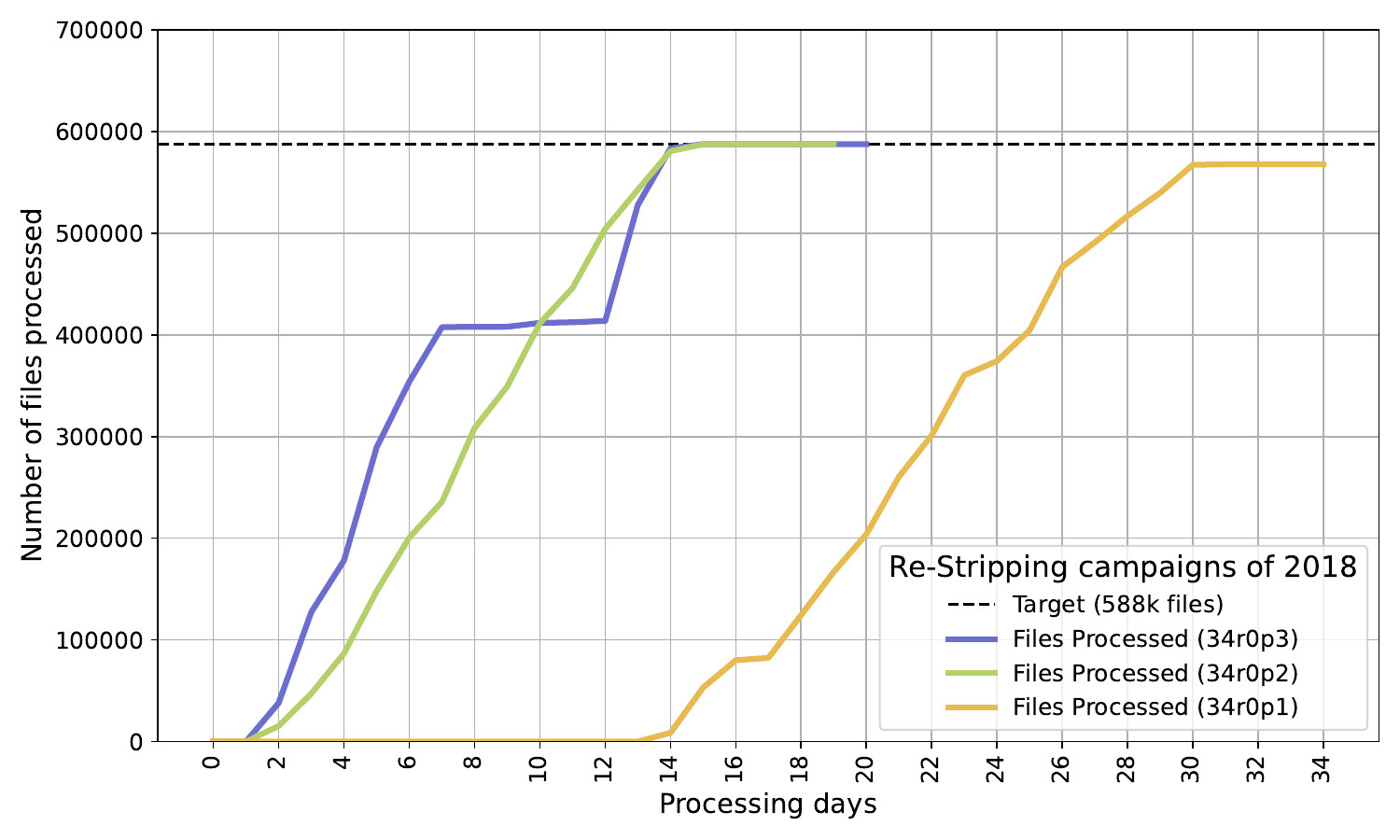}
    \caption{Comparison of several campaigns processing the 2018 data set, plotting versus the number of days of processing rather than the actual calendar dates.
    The curve for the campaign (Stripping)34r0p1 does not reach the "target line" as the other two campaigns because, at that time, certain runs or files had been marked as unusable for physics based on data quality assessments. The curve does not otherwise reflect any operational issue.  The multi-day pause in the middle of Stripping34r0p3 was due to planned resource inavailability from the distributed computing team, and is not considered in the overall production time comparison.}
    \label{fig:comparison_campaigns_Data2018_vs_nr_days}
\end{figure}

Figure~\ref{fig:comparison_campaigns_Data2018_vs_nr_days} shows the final three incremental re-Stripping campaigns performed on the 2018 data sample. Each campaign processing time showed improvement with each subsequent version, with the first reprocessing time taking two weeks, and the latest taking one week after removing scheduled operational pauses.

While legacy data software stacks often lag behind modern HEP developments, reprocessing older data with contemporary workflows can help meet experimental constraints. Analyses following major Stripping campaigns are crucial for procedural improvements, particularly given the high participant turnover in sporadic legacy productions. Targeted training programs have proven effective, as demonstrated by recent surveys showing enhanced participant confidence and professional development. Simultaneously, workflow optimization and progress monitoring are essential for maintaining operational efficiency in large reprocessing campaigns. Together, these human and technical factors enable more effective legacy data production while supporting researcher development.

\section{Summary}

Building on the successful Run-3 Sprucing model, DPA WP5 proposes an iterative workflow for developing new Stripping campaigns for Runs 1 and 2 data. The approach features: (1) early line development when physics ideas emerge, (2) branch-based analyst work with CI testing, (3) regular PWG branch integration, (4) milestone tracking for resource planning, and (5) buffered deadlines for alignment with PPG and OPG priorities. This flexible system allows analysts to develop when convenient while providing coordination teams evidence for campaign planning, as demonstrated by the successful most-recent campaign carried out in late 2023 and early 2024 aforementioned.

The LHCb collaboration maintains a vibrant legacy program using Runs 1 and 2 datasets through sustainable systems combining software development and large-scale reprocessing. The Stripping framework's Python-based flexibility, combined with modern tools and strong computing and operations collaboration, manages campaign complexity while offering development opportunities for researchers at all career stages, even as focus shifts to Run 3 data.

\backmatter


\bmhead{Acknowledgements}

The authors would like to greatly thank the myriad of analysts who have contributed to the Stripping project from LHCb over decade of application.  Additionally, we thank the significant support of the LHCb Computing project for maintaining and developing the core software upon which the Stripping project relies.  Finally, we thank the LHCb production team for running the processing campaigns.

The work of NG is supported by the US National Science Foundation through award PHY-2411665. NSa would like to acknowledge support from the UK Science and Technology Facilities Council (STFC). 









        








\begin{appendices}


\end{appendices}

\bibliography{sn-bibliography}

\end{document}